\documentclass[12pt]{iopart}
\usepackage[]{latexsym}
\newcommand{\be}{\begin{eqnarray}}
\newcommand{\ee}{\end{eqnarray}}

\begin{document}
\title{Newton's law in an effective non commutative space-time}
\author{Alessandro Gruppuso}

\address{Dipartimento di Fisica, Universit\`a di Bologna and INFN,
Sezione di Bologna \\
via Irnerio 46, 40126, Bologna, Italy
\\ and \\
\footnote[1]{Present address.}INAF/IASF, Istituto di Astrofisica Spaziale e Fisica Cosmica \\
Sezione di Bologna\\
via Gobetti 101, 40129, Bologna, Italy}

\ead{gruppuso@bo.iasf.cnr.it}

\begin{abstract}

The Newtonian Potential is computed exactly  in a theory that is
fundamentally Non Commutative in the space-time coordinates.
When the dispersion for the distribution of the source is
minimal (i.e. it is equal to the non commutative parameter $\theta$),
the behavior for large and small distances is analyzed.

\end{abstract}

\pacno{11.10.Nx}


\maketitle
\raggedbottom
\setcounter{page}{1}
%
%

%
%
{\it Introduction}. Recently a new model of Quantum Field Theory on Non-Commutative
space time, satisfying Lorentz Invariance and Unitarity,
has been proposed in \cite{AnaisEuro,AnaisEuro2}: it is shown that
there is no need of Weyl-Wigner-Moyal $\star$-product \cite{weyl,wigner,moyal} and the
Non-Commutativity is carried by a gaussian cut-off in the Fourier
Transform of the fields. This is not an ad hoc regularization
device but is a result coming from the averaging operation on
coherent states. This cut-off (which depends on the Non-Commutative
parameter $\theta $ \cite{AnaisEuro2}) is also present in the (Feynman) Propagator
and is responsible for the UV finiteness of the theory
\cite{AnaisEuro,AnaisEuro2,Nicolini}.

The aim of this paper is to compute the modification of the
Newtonian Potential due to the change of the Green's Function
caused by the Non-Commutativity of the space-time coordinates.


%
%
%
%

\vskip 1 truecm

In order to deal with a theory that is effectively Non-Commutative
in the space-time coordinates, we follow \cite{AnaisEuro} and
suppose that the propagator satisfies: \be -\Box _{x}
G(x,x^{\prime}) = \left(\sqrt{{a\over \pi}}\right)^4 e^{-a (x -
x^{\prime})^2} \, , \label{defprop} \ee where $\Box_{x}=\partial_t
^2 + \nabla_x ^2$, $ a = 1/ 4 \theta$ and $\theta $ is the
Non-Commutative parameter with the dimensions of an area \footnote{We
are considering an euclidean signature of the metric, coming
from a Wick rotation of the time coordinate.}. Notice that in the
limit $\theta \rightarrow 0 $ the gaussian distribution becomes a
delta function and standard commutative theory is recovered.
As usual the Newtonian Potential $V(x)$ is related to the
fluctuation of the $00$ component of the metric $h_{00}(x)$
through $V(x) = h_{00}(x) /2 $ and the fluctuation is given by
\cite{wald}: \be h_{00} (x) = - 8 \pi G_{N} \int d^4 x^{\prime}
G(x,x^{\prime}) T_{00}(x^{\prime}) \, , \label{defh00} \ee where
$G_{N}$ is the Newton's constant and $T_{00}$ is the $00$
component of the stress-tensor of the source of the gravitational
field. Since in a theory that is fundamentally Non-Commutative the
concept of point is meaningless, the source cannot be a delta
function but will be given by a gaussian distribution: \be
 T_{00}(x) = \int {d^3 q \over {(2 \pi)^3}} e^{-i \vec{q} \cdot \vec{x}} f(q^2)
 \, ,
 \label{T00}
\ee where $f(q^2) = e^{-\alpha \vec{q}^2} M $, with $\alpha $ and
$M $ constants. We shall choose $\alpha = \theta $ because we want
to consider a minimum dispersion for the source.

%
%
%
%
%
%
%
%

The Fourier Transform of the propagator is called $G(k)$ and it is implicitly defined as
\be
G(x,x^{\prime}) =\int {d^4 k \over {(2 \pi)^4}} e^{i k_{\mu} (x^{\mu} - x^{\prime \mu})} G(k)
\, .
\label{G(k)}
\ee
Considering that the Fourier Transform of a gaussian distribution
is still a gaussian function, \be \left(\sqrt{{1\over
{4 \theta \pi}}}\right)^4 e^{-{1 \over{4 \theta}} (x -
x^{\prime})^2} = \int {d^4 k \over {(2 \pi)^4}} e^{i
k_{\mu} (x - x^{\prime})^{\mu}} e^{-\theta(k_0^2 + \vec{k}^2)} \,
,
\ee then $G(k)$ can be obtained through
equation (\ref{defprop}): \be G(k) = {e^{-\theta(k_0^2 + \vec{k}^2)}
\over {k_0 ^2 + \vec{k}^2}} \; . \label{propgk} \ee

%
%
%
From equations (\ref{defh00}),(\ref{T00}),(\ref{G(k)}) and
(\ref{propgk}) one can write
\be \fl h_{00}(x) = - 8 \pi
G_{N} \int d^4 x^{\prime} \int {d^4 k \over {(2 \pi)^4}} e^{i
k_{\mu} (x^{\mu} - x^{\prime \mu})}{e^{-\theta( k_0 ^2 +\vec{k}^2)} \over {k_0 ^2 +
\vec{k}^2}}
 \int {d^3 q \over {(2 \pi)^3}} e^{-i \vec{q} \cdot \vec{x}^{\prime}} f(q^2)
 \, .
 \label{h00uno}
\ee Since the source does not depend on time we can use \be \int
dt^{\prime} e^{-i k_0 t^{\prime}} = 2 \pi \delta(k_0) \, ,
\label{dk0} \ee and with the help of \be \int d^3x^{\prime} e^{i
\vec{x}^{\prime}\cdot (\vec{k} -\vec{q})} = (2 \pi)^3
\delta(\vec{k}-\vec{q}) \, , \label{dkq} \ee it is possible to
simplify (\ref{h00uno}), obtaining the following equation: \be
h_{00}(x) = - 8 \pi G_{N}  \int {d^3 k \over {(2 \pi)^3}} e^{i
\vec{k}\cdot \vec{x}}
 {e^{-\theta \vec{k}^2} \over { \vec{k}^2}} f(k^2)
 \, .
 \label{h00due}
\ee
Performing the integration over the angular part of $d^3 \, k$ one finds
\be
h_{00}(r) = - {2 \, G_{N} \over { \pi }} {M \over {i r}}
\int_{-\infty}^{\infty} dk \, k
e^{i k r }
 {e^{-(\theta + \alpha) k^2} \over { k^2 + \epsilon ^2}}
 \, ,
 \label{h00tre}
\ee where the definition of $f(k^2)$ has been used, with $k =
\sqrt{ {\vec k}^2 }$, and the limit $\epsilon \rightarrow 0$
is understood \footnote{As usual this limit has been introduced in
order to regularize the integrand at $k^2=0$.}. The integration can
be performed exactly \cite{grad} and the
Newtonian Potential is found to be the following:
\be V(r) = - G_N {M \over r}
Erf\left({r \over{ 2 \sqrt{\theta + \alpha }}}\right) \, ,
\label{npotential}
\ee
where $V(r) =  h_{00}(r)/2 $ has been used.


%
%


First of all we notice that if $ \alpha \gg \theta $ then the
Non Commutativity is screened by the source.
This is natural since the source must be localized as much as possible
in space. But in a theory that is fundamentally Non-Commutative,
the smallest spread is given by $\theta$.

We set now $\alpha = \theta $ since we wish to analyze the
minimum case. In the region for which $r \gg 2\sqrt{2 \, \theta}$
one obtains: \be V(r) = - G_N {M \over r} \left[ 1 + e^{-r^2 / (8 \,\theta)} \left(
-{2\over { \sqrt{\pi}}} {\sqrt{2 \, \theta } \over r} + {\cal O}
\left( {2 \sqrt{2 \theta} \over{ r }}\right)^3 \right) \right] \,
.
\ee Numerically (on requiring that the correction is
of order $1$) one finds that the large distance correction
becomes important for the following critical value \be {r_c \over
{2 \sqrt{2 \theta}}} \simeq 0.4576 \, . \ee Since it is known that
Newton's Law is verified up to a distance of the order of $200 \,
\mu$m \cite{experiment}, it is then possible to find a constraint
for $\theta$, $ \theta < 10^{-8} m^2 \, $. Of course due to the
fact that Newton's Law is not tested for very small
distances, this constraint is not so strong. One has to use
precision measurements in order to have a significant bound. In
the opposite region where $r \ll 2\, \sqrt{2 \, \theta}$ one
obtains \be V(r) = - G_N {M \over r} \left[  {r \over {\sqrt{\pi(2
\, \theta)}}} + {\cal O}\left( {r \over{ 2 \sqrt{2 \theta}
}}\right)^3 \right] \, . \ee It is interesting to note that
there is no divergence (unlike the Commutative case): the
introduction of the $\theta$ parameter gives a minimal length that
regularizes the theory: \be V(0) = - G_N {M \over
{\sqrt{\pi(2 \,\theta)}}} \, . \ee



%
%
%

\vskip 1 truecm

{\it Conclusions}. The main result of this brief paper is given by equation (\ref{npotential}).
The calculation is exact and no approximation has been made.
We notice that there is a deviation from the standard Newtonian
Potential but the shape of the source can screen the effect of
Non-Commutative space. In the minimal case (i.e. the dispersion
for the profile of the source is given by $\theta$) such a
deviation has been computed for the large and small distance
approximations. In the first case we used the corrections to
Newton's Law in order to constraint the Non-Commutative parameter 
(but the obtained bound is weak since Newton's Law is
verified down to distances of the order of $200 \, \mu$m). In the
second case it is seen that there is a regular behavior at $r=0$.
This reflects the well known fact that Non-Commutativity
introduces a minimal length which regularizes the theory.


%

\vskip 1 truecm

{\it Aknoledgements}. It is a pleasure to thank Euro Spallucci for useful suggestions
and many comments on the draft version of this paper. I am also
grateful to Piero Nicolini, Anais Smailagic and Giovanni Venturi for interesting
discussions. I thank Roberto Casadio for collaboration in the
early stage of this work.



%
%
%

\vskip 1 truecm

\section*{References}
\end{document}